\documentstyle[eqsecnum,multicol,aps,prl,epsf]{revtex}

\date{\today}
\draft
\tighten
\begin{document}
\def\sqr#1#2{{\vcenter{\hrule height.3pt
      \hbox{\vrule width.3pt height#2pt  \kern#1pt
         \vrule width.3pt}  \hrule height.3pt}}}
\def\square{\mathchoice{\sqr67\,}{\sqr67\,}\sqr{3}{3.5}\sqr{3}{3.5}}
\def\today{\ifcase\month\or
  January\or February\or March\or April\or May\or June\or July\or
  August\or September\or October\or November\or December\fi
  \space\number\day, \number\year}

\def\Bbb{\bf}

\renewcommand{\a}{\alpha}
\renewcommand{\b}{\beta}
\newcommand{\g}{\gamma}           \newcommand{\G}{\Gamma}
\renewcommand{\d}{\delta}         \newcommand{\D}{\Delta}
\newcommand{\ve}{\varepsilon}
\newcommand{\eps}{\epsilon}
\newcommand{\k}{\kappa}
\newcommand{\ld}{\lambda}        \newcommand{\LD}{\Lambda}
\newcommand{\om}{\omega}         \newcommand{\OM}{\Omega}
\newcommand{\p}{\psi}             \newcommand{\PS}{\Psi}
\newcommand{\ro}{\rho}
\newcommand{\s}{\sigma}           \renewcommand{\S}{\Sigma}
\newcommand{\th}{\theta}         \newcommand{\T}{\Theta}
\newcommand{\f}{{\phi}}           \newcommand{\F}{{\Phi}}
\newcommand{\vf}{{\varphi}}
\newcommand{\y}{{\upsilon}}       \newcommand{\Y}{{\Upsilon}}
\newcommand{\z}{\zeta}
\newcommand{\X}{\Xi}
\newcommand{\cA}{{\cal A}}
\newcommand{\cB}{{\cal B}}
\newcommand{\cC}{{\cal C}}
\newcommand{\cD}{{\cal D}}
\newcommand{\cE}{{\cal E}}
\newcommand{\cF}{{\cal F}}
\newcommand{\cG}{{\cal G}}
\newcommand{\cH}{{\cal H}}
\newcommand{\cI}{{\cal I}}
\newcommand{\cJ}{{\cal J}}
\newcommand{\cK}{{\cal K}}
\newcommand{\cL}{{\cal L}}
\newcommand{\cM}{{\cal M}}
\newcommand{\cN}{{\cal N}}
\newcommand{\cO}{{\cal O}}
\newcommand{\cP}{{\cal P}}
\newcommand{\cQ}{{\cal Q}}
\newcommand{\cS}{{\cal S}}
\newcommand{\cR}{{\cal R}}
\newcommand{\cT}{{\cal T}}
\newcommand{\cU}{{\cal U}}
\newcommand{\cV}{{\cal V}}
\newcommand{\cW}{{\cal W}}
\newcommand{\cX}{{\cal X}}
\newcommand{\cY}{{\cal Y}}
\newcommand{\cZ}{{\cal Z}}
\newcommand{\hA}{{\widehat A}}
\newcommand{\hB}{{\widehat B}}
\newcommand{\hC}{{\widehat C}}
\newcommand{\hD}{{\widehat D}}
\newcommand{\hE}{{\widehat E}}
\newcommand{\hF}{{\widehat F}}
\newcommand{\hG}{{\widehat G}}
\newcommand{\hH}{{\widehat H}}
\newcommand{\hI}{{\widehat I}}
\newcommand{\hJ}{{\widehat J}}
\newcommand{\hK}{{\widehat K}}
\newcommand{\hL}{{\widehat L}}
\newcommand{\hM}{{\widehat M}}
\newcommand{\hN}{{\widehat N}}
\newcommand{\hO}{{\widehat O}}
\newcommand{\hP}{{\widehat P}}
\newcommand{\hQ}{{\widehat Q}}
\newcommand{\hS}{{\widehat S}}
\newcommand{\hR}{{\widehat R}}
\newcommand{\hT}{{\widehat T}}
\newcommand{\hU}{{\widehat U}}
\newcommand{\hV}{{\widehat V}}
\newcommand{\hW}{{\widehat W}}
\newcommand{\hX}{{\widehat X}}
\newcommand{\hY}{{\widehat Y}}
\newcommand{\hZ}{{\widehat Z}}
\newcommand{\Ha}{{\widehat a}}
\newcommand{\Hb}{{\widehat b}}
\newcommand{\Hc}{{\widehat c}}
\newcommand{\Hd}{{\widehat d}}
\newcommand{\He}{{\widehat e}}
\newcommand{\Hf}{{\widehat f}}
\newcommand{\Hg}{{\widehat g}}
\newcommand{\Hh}{{\widehat h}}
\newcommand{\Hi}{{\widehat i}}
\newcommand{\Hj}{{\widehat j}}
\newcommand{\Hk}{{\widehat k}}
\newcommand{\Hl}{{\widehat l}}
\newcommand{\Hm}{{\widehat m}}
\newcommand{\Hn}{{\widehat n}}
\newcommand{\Ho}{{\widehat o}}
\newcommand{\Hp}{{\widehat p}}
\newcommand{\Hq}{{\widehat q}}
\newcommand{\Hs}{{\widehat s}}
\newcommand{\Hr}{{\widehat r}}
\newcommand{\Ht}{{\widehat t}}
\newcommand{\Hu}{{\widehat u}}
\newcommand{\Hv}{{\widehat v}}
\newcommand{\Hw}{{\widehat w}}
\newcommand{\Hx}{{\widehat x}}
\newcommand{\Hy}{{\widehat y}}
\newcommand{\Hz}{{\widehat z}}
\newcommand{\deff}{\,\stackrel{\rm def}{\equiv}\,}
\newcommand{\lra}{\longrightarrow}
\newcommand{\ra}{\,\rightarrow\,}
\def\limar#1#2{\,\raise0.3ex\hbox{$\longrightarrow$\kern-1.5em\raise-1.1ex
\hbox{$\scriptstyle{#1\rightarrow #2}$}}\,}
\def\limarr#1#2{\,\raise0.3ex\hbox{$\longrightarrow$\kern-1.5em\raise-1.3ex
\hbox{$\scriptstyle{#1\rightarrow #2}$}}\,}
\def\limlar#1#2{\ \raise0.3ex
\hbox{$-\hspace{-0.5em}-\hspace{-0.5em}-\hspace{-0.5em}
\longrightarrow$\kern-2.7em\raise-1.1ex
\hbox{$\scriptstyle{#1\rightarrow #2}$}}\ \ }
\newcommand{\limm}[2]{\lim_{\stackrel{\scriptstyle #1}{\scriptstyle #2}}}
\newcommand{\wt}{\widetilde}
\newcommand{\os}{{\otimes}}
\newcommand{\da}{{\dagger}}
\newcommand{\stimes}{\times\hspace{-1.1 em}\supset}
\def\h{\hbar}
\newcommand{\ih}{\frac{\i}{\h}}
\newcommand{\exx}[1]{\exp\left\{ {#1}\right\}}
\newcommand{\ord}[1]{\mbox{\boldmath{$\cO$}}\left({#1}\right)}
\newcommand{\one}{{\leavevmode{\rm 1\mkern -5.4mu I}}}
\newcommand{\Z}{Z\!\!\!Z}
%
\newcommand{\Ibb}[1]{ {\rm I\ifmmode\mkern
            -3.6mu\else\kern -.2em\fi#1}}
\newcommand{\ibb}[1]{\leavevmode\hbox{\kern.3em\vrule
     height 1.2ex depth -.3ex width .2pt\kern-.3em\rm#1}}
\newcommand{\N}{{\Ibb N}}
\newcommand{\C}{{\ibb C}}
\newcommand{\R}{{\Ibb R}}
\newcommand{\HH}{{\Ibb H}}
\newcommand{\rational}{{\kern .1em {\raise .47ex
\hbox{$\scripscriptstyle |$}}
    \kern -.35em {\rm Q}}}
\newcommand{\bm}[1]{\mbox{\boldmath${#1}$}}
\newcommand{\intf}{\int_{-\infty}^{\infty}\,}
\newcommand{\LL}{\cL^2(\R^2)}
\newcommand{\LLS}{\cL^2(S)}
\newcommand{\Ree}{{\cal R}\!e \,}
\newcommand{\Imm}{{\cal I}\!m \,}
\newcommand{\tr}{{\rm {Tr} \,}}
\newcommand{\er}{{\rm{e}}}
\renewcommand{\i}{{\rm{i}}}
\newcommand{\divv}{{\rm {div} \,}}
\newcommand{\id}{{\rm{id}\,}}
\newcommand{\ad}{{\rm{ad}\,}}
\newcommand{\Ad}{{\rm{Ad}\,}}
\newcommand{\const}{{\rm{\,const\,}}}
\newcommand{\rank}{{\rm{\,rank\,}}}
\newcommand{\diag}{{\rm{\,diag\,}}}
\newcommand{\sign}{{\rm{\,sign\,}}}
\newcommand{\pa}{\partial}
\newcommand{\pad}[2]{{\frac{\partial #1}{\partial #2}}}
\newcommand{\padd}[2]{{\frac{\partial^2 #1}{\partial {#2}^2}}}
\newcommand{\paddd}[3]{{\frac{\partial^2 #1}{\partial {#2}\partial {#3}}}}
\newcommand{\der}[2]{{\frac{{\rm d} #1}{{\rm d} #2}}}
\newcommand{\derr}[2]{{\frac{{\rm d}^2 #1}{{\rm d} {#2}^2}}}
\newcommand{\fud}[2]{{\frac{\delta #1}{\delta #2}}}
\newcommand{\fudd}[2]{{\frac{\d^2 #1}{\d {#2}^2}}}
\newcommand{\fuddd}[3]{{\frac{\d^2 #1}{\d {#2}\d {#3}}}}
\newcommand{\dpad}[2]{{\displaystyle{\frac{\partial #1}{\partial #2}}}}
\newcommand{\dfud}[2]{{\displaystyle{\frac{\delta #1}{\delta #2}}}}
\newcommand{\dd}{\partial^{(\ve)}}
\newcommand{\ddd}{\bar{\partial}^{(\ve)}}
\newcommand{\dfrac}[2]{{\displaystyle{\frac{#1}{#2}}}}
\newcommand{\dsum}[2]{\displaystyle{\sum_{#1}^{#2}}}
\newcommand{\dint}{\displaystyle{\int}}
\newcommand{\dg}{\!\not\!\partial}
\newcommand{\vg}[1]{\!\not\!#1}
\def\<{\langle}
\def\>{\rangle}
\def\lgl{\langle\langle}
\def\rgr{\rangle\rangle}
\newcommand{\bra}[1]{\left\langle {#1}\right|}
\newcommand{\ket}[1]{\left| {#1}\right\rangle}
\newcommand{\vev}[1]{\left\langle {#1}\right\rangle}
\newcommand{\be}{\begin{equation}}
\newcommand{\ee}{\end{equation}}
\newcommand{\bn}{\begin{eqnarray}}
\newcommand{\en}{\end{eqnarray}}
\newcommand{\bnn}{\begin{eqnarray*}}
\newcommand{\enn}{\end{eqnarray*}}
\newcommand{\e}{\label}
\newcommand{\nbr}{\nonumber\\[2mm]}
\newcommand{\r}[1]{(\ref{#1})}
\newcommand{\refp}[1]{\ref{#1}, page~\pageref{#1}}
\renewcommand {\theequation}{\thesection.\arabic{equation}}
\renewcommand {\thefootnote}{\fnsymbol{footnote}}
\newcommand{\qq}{\qquad}
\newcommand{\qqq}{\quad\quad}
\newcommand{\biz}{\begin{itemize}}
\newcommand{\eiz}{\end{itemize}}
\newcommand{\ben}{\begin{enumerate}}
\newcommand{\een}{\end{enumerate}}

\def\rep{representation}
\def\nc{noncommutative }
\def\ncy{noncommutativity }
\def\com{commutative }
\def\sm{Standard Model }
\def\Uo{$U_{\star}(1)$ }
\def\Ut{$U_{\star}(2)$ }
\def\Uth{$U_{\star}(3)$ }
\def\Un{$U_{\star}(n)$}
\def\uo{$u_{\star}(1)$ }
\def\ut{$u_{\star}(2)$ }
\def\uth{$u_{\star}(3)$}
\def\un{$u_{\star}(n)$}
\def\sp{$\star$-product} 
\def\um{$u_{\star}(m)$}
\def\Um{$U_{\star}(m)$}
\def\nbyn {$n\!\times\! n$}

\title{Noncommutative Gauge Field Theories:\\
A No-Go Theorem}

\author{M. Chaichian$^{\dagger}$,
P. Pre\v{s}najder$^{\dagger,a}$, M. M. Sheikh-Jabbari$^{\dagger,b,c}$
\ \ and \ \ {{A. Tureanu}}$^{\dagger}$}

\address{$^{\dagger}$High Energy Physics Division, Department of
Physics,
University of Helsinki\\
\ \ {and}\\
\ \ Helsinki Institute of Physics,
P.O. Box 64, FIN-00014 Helsinki, Finland\\
$^a$Department of Theoretical Physics, Comenius University, Mlynsk\'{a} dolina, SK-84248 Bratislava,
Slovakia \\
$^{b}$ The Abdus Salam ICTP,
Strada Costiera 11,Trieste, Italy\\
$^c$Institute for Studies in Theoretical Physics and Mathematics\\
P.O. Box 19395-5531, Tehran, Iran}

\maketitle
\setcounter{footnote}{0}

\begin{abstract} 

Studying the general structure of the \nc (NC) local groups, we prove a no-go theorem for NC 
gauge theories. According to this theorem, the closure condition of the gauge algebra
implies that:  1) the local NC $u(n)$ {\it algebra} only admits
the irreducible \nbyn\ matrix-representation. Hence the gauge fields are in \nbyn\ matrix form, 
while the matter fields {\it can only
be} in fundamental, adjoint or singlet states;
2) for any gauge group consisting of several
simple-group factors, the matter fields can transform nontrivially under
{\it at most two} NC group factors. In other words, the matter fields cannot 
carry more than two NC gauge group charges. This no-go theorem
imposes strong restrictions on the NC version of the
Standard Model and in resolving the standing problem of charge
quantization in \nc QED.

\end{abstract}

\pacs{PACS: 11.15.-q, 11.30.Er, 11.25.Sq.
\hspace {1cm} HIP-2001-24/TH, IPM/P-2001/017, hep-th/0107037}
\begin{multicols} {2}

\section{Introduction}
\setcounter{equation}{0}

During the past two years, there has been a lot of work devoted
to the theories formulated on the 
\nc space-time \cite{SW}. Apart from the string theory interests, the
field theories on
\nc space-times (Moyal plane)
have their own attractions.  To obtain a \nc version of the action for
any given field theory one 
should replace the 
usual product of the functions (fields) with the $\star$-product: 
\begin{eqnarray}\label{star}
(f\star g)(x)&=&\exx{{\i\over 2}\theta_{\mu\nu}
\partial_{x_{\mu}}\partial_{y_{\nu}}}f(x)g(y)\Big|_{x=y}\cr
&=&f(x)g(x)+{\i\over2}\theta_{\mu\nu}
\partial_{{\mu}}f\partial_{{\nu}}g+\ord{\theta^2}\ ,
\end{eqnarray}
where $\theta_{\mu\nu}=-\theta_{\nu\mu}$. 
In a more mathematical way, the fields of a \nc field theory should be chosen
from the C$^{\star}$-algebra of
functions with the above \sp. 

Although in the \nc case the Lorentz symmetry is explicitly broken, one
can still realize the same representations as in the commutative case
where, depending on the space-time (Lorentz) representation, the fields
can be scalars, Dirac fields, vector bosons, etc. This can be done noting
the notion of "trace" in the corresponding
C$^{\star}$-algebra. This "trace" is basically the integration
over the whole space-time, which is already there in the usual definition
of the action. Therefore, the notion of "trace" implies that the \sp\ in the
quadratic terms of the
actions can be removed. In other words, only the interaction terms in the
action receive some corrections due to the \sp\
\cite{{Filk},{CDP1},{Shren},{Thomas}}.  
However, in order to
quantize the theory, besides the quadratic parts of the action, we should
specify the Hilbert space of the theory (or, equivalently, the measure in
the path integral quantization), as well as the conjugate momentum of any
field. For the space-like noncommutativity, $\theta_{\mu\nu}\theta^{\mu\nu}>0$, this
Hilbert space (or the path integral measure) can consistently be chosen
the same as in the commutative case \cite{{uni1},{uni2},{masud}}.

The next step in constructing a physical \nc model is to develop the
concept of local gauge symmetry. Intuitively, because of the inherent
non-locality induced by the \sp\ (\ref{star}), the notion of {\it local}
symmetry in the \nc case should be handled with special care. As a result,
the pure \nc $U(1)$ theory, which we shall denote by $U_{\star}(1)$,
behaves similarly to the usual non-Abelian gauge theories, but now the
structure constants depend on the momenta of the fields \cite{Shren}. We shall
discuss this point in more detail later.

However, before turning to more physical questions, one should develop the
\nc groups underlying the gauge theories, as well as their
representations. In general, as discussed in \cite{{NCSO},{Bars}}, it is not trivial to
define the \nc version of usual simple local groups, as the \sp\ will
destroy the closure condition. For example, let $g_1$ and $g_2$ be two
traceless hermitian $x$-dependent $n\times n$ matrices (elements of the
usual $su(n)$). It is very easy to check that $g_1\star g_2 - g_2\star
g_1$ is not traceless anymore. Consequently, the only group which admits a
simple \nc extension is $U(n)$ (we denote this extension by \Un). The \nc extensions of the
other groups are not trivially obtained by the insertion of the \sp. However, the
\nc $SO$ and $USp$ algebras have been constructed in a more involved way
\cite{{NCSO},{Bars}}.

Besides the simple \nc group \Un\ and its representations, we also need to
define the direct product of \nc groups, such as \Un$\times$\Um.
In this case, since the group elements are matrix valued
functions, in general the usual definition of the direct product of
group elements does not work. We will show that this fact imposes
strong restrictions on the matter fields (those which are in fundamental
representations).

This work is organized as follows. In the next section, reviewing the pure
\nc Yang-Mills theories, we discuss the gauge invariance issue in more
detail and show that the only
possible \rep\ for the \un\ algebra is the one generated through \nbyn\ hermitian matrices.
In section 3, we consider the matter fields and their gauge invariance. 
This will lead to the charge quantization for \nc QED \cite{Haya}. Then we
proceed with the cases in which our gauge group consists of a direct
product of some simple \Un\ factors. We show that group theory
considerations (closure condition) will restrict our matter fields to
be charged at most under {\it two} \Un\ factors of our gauge group.  
We close this work with discussions and conclusions.
  
\vspace{-.5cm}
\section{Pure $U_{\star}(n)$ gauge theories}
\setcounter{equation}{0}
\vspace{-.3cm}  
To define the pure \Un\ Yang-Mills theory, we start with introducing the \Un\ group and
the corresponding algebra. The \un\ algebra is generated by $n\times n$ hermitian matrices
whose elements (which are complex valued functions) are multiplied by the \sp\
(\ref{star}) \cite{Armoni}. 
If we denote the usual $n\times n$ $su(n)$ generators by $T^a,\ a=1,2,\cdots, n^2-1$, normalized
 as $Tr(T^aT^b)= {1\over 2} \delta^{ab}$, by adding 
$T^0={1\over\sqrt{2n}}{\bf 1}_{n\times n}$ we can cover all $n\times n$ hermitian
matrices\footnote{We note that the normalization factor ${1\over\sqrt{2n}}$ for $T^0$ is chosen
conveniently, so that $d^{ABC}={2} Tr(\{T^A, T^B\} T^C)$ is totally symmetric. As it
is expected, the renormalizability of the gauge theory does not depend on the relative
normalization for the $su(n)$ and $u(1)$ generators. We are grateful to L. Bonora and
M. Salizzoni for a discussion on this point.} \cite{Loriano}. 
Then any element of \un\ can be expanded as
\be\label{un-algebra}
f=\sum_{A=0}^{n^2-1}\ f^A(x) T^A\ , 
\ee
and the \un\ Lie-algebra is defined with the star-matrix bracket:
\be\label{un-bracket}
[f, g]_{\star}= f\star g - g \star f\; ,\;\;\;\;\;\  f,g\in u_{\star}(n)\ .
\ee 
Evidently the above bracket closes on the \un\ algebra. For the case of $n=1$, the \uo
case, the above bracket reduces to the so-called Moyal bracket.

The \Un\ gauge theory is described by the \un\ valued gauge fields
\be\label{un-g.f.}
G_{\mu}=\sum_{A=0}^{n^2-1}\ G_{\mu}^A(x) T^A\ .
\ee
It is straightforward to show that the field strength
\be\label{f-s}
G_{\mu\nu}=\partial_{[\mu}G_{\nu]} + i g[G_{\mu},G_{\nu}]_{\star}\ ,
\ee
under the infinitesimal \un\ gauge transformations:
\be\label{g.t.}
G_{\mu}\to\ G'_{\mu}= G_{\mu} + i\partial_{\mu} \lambda + g [\lambda, G_\mu]_{\star}\ ,
\; \lambda\in u_{\star}(n)\ , 
\ee 
transforms covariantly:
\be 
G_{\mu\nu}\to\ G'_{\mu\nu}= G_{\mu\nu} +i g [\lambda,G_{\mu\nu}]_{\star}\ .
\ee
To construct the gauge invariant action we need to define a "trace" in the C$^\star$-algebra
of the functions (elements of $n\times n$ matrices). It can be shown that the integration over
the space-time can play the role of this trace; it enjoys the cyclic permutation symmetry
and can be normalized. Hence, the action we are looking for is
\be\label{G-action}
S=-{1\over 4\pi}\int d^Dx\ {\rm Tr}(G_{\mu\nu}\star G^{\mu\nu})\ ,
\ee
where the trace is taken over the $n\times n$ matrices.

The first peculiar feature of the pure \Un\ gauge theory we would like to mention here is
that, fixing the number of gauge field degrees of freedom (which is $n^2$) the dimension
of
the matrix \rep\ is automatically fixed, i.e. the gauge fields must be in the \nbyn\ matrix
form. This is a specific
property dictated by \ncy and
in particular the fact that the algebra bracket (\ref{un-bracket}) also involves the \sp.
To make it clear, let us consider a particular example of \ut and take the $3\!\times\! 3$
\rep\ for the matrix part, which we denote by $\Sigma^i, i=1,2,3$ and ${\bf 1}_{3\!\times\!
3}$. It is easy to see  that in order to close the algebra with the star-matrix bracket
(\ref{un-bracket}), in fact besides the $\Sigma^i$'s we need all the other six
$3\times 3$ hermitian matrices. Therefore, the algebra is not \ut anymore (it is what we call
\uth). The above argument for \ut\ can be generalized to the \un\ case. Let us
start with an irreducible $N\!\times\! N$ \rep\ ($N\!\geq\! n$).
The enveloping algebra of $u(n)$ for this \rep\ closes in $u(N)$ (and not $u(n)$), unless
$N=n$ or otherwise our \rep\ is reducible. Therefore, this irreducible $N\!\times\! N$ \rep\
($N\!>\! n$)
is not forming a proper basis for \un\ gauge fields. 

The finite \Un\ gauge transformations are generated by the elements of the group (in the
adjoint \rep) which are obtained by star-exponentiation of the elements of the algebra:
\be
U=({\rm e}\star)^{i\lambda} = 1+i\lambda -{1\over 2} \lambda\star\lambda 
\cdots\ , \;\;\ U\in U_{\star}(n)\ .
\ee   
Then under finite gauge transformations $G_\mu$ should transform as
\be
G_{\mu}\to\ G'_{\mu}= U\star G_{\mu}\star U^{-1}  + {i\over g} U\star\partial_{\mu}U^{-1}\ .
\ee
It can be easily checked that $G_{\mu\nu}\to U\star G_{\mu\nu} \star U^{-1}$ and hence the
action (\ref{G-action}) remains invariant.
  
\vspace{-.5cm}
\section{Matter fields}
\setcounter{equation}{0}
\vspace{-.4cm}
Now that we have introduced the pure \Un\  gauge theory and the adjoint \rep\ 
of \Un\ group we are ready to add the matter fields, which are in the fundamental \rep\ of
the group. Hence if we denote the matter fields by $\psi$, under gauge transformations
\cite{Haya}
\be\label{psi g.t.}
\psi\to \psi'=U\star\psi\ .
\ee
Of course, the anti-fundamental \rep\ is also possible:
\be\label{chi g.t.}
\chi\to \chi'=\chi\star U^{-1}\ .
\ee

For the fermionic (Dirac) matter fields, it is straightforward to show that 
the action for the $\psi$-field, 
defined as \cite{{Haya}}
\bn\label{psi-action}
S_f &=& \int d^Dx\ {\bar\psi}\gamma^{\mu}D_{\mu}\star\psi\ , \cr
D_\mu &=& \partial_{\mu}+i g G_{\mu}\ ,
\en
is invariant under the above gauge transformations. We also note that $\psi$ and the
anti-fundamental matter field, $\chi$, are related by the  \nc version of charge
conjugation \cite{CPT}.
\begin{center}
{\bf A. Charge quantization}
\end{center}

Before proceeding with the more complicated gauge groups we would like to point out a
peculiar property of the \Uo\ theory with matter fields, which may be called NCQED.   
It is well known that in the non-Abelian gauge theories the corresponding "charge" is
fixed by specifying the \rep\ of the fields (like the $SU(2)$ weak charges in the usual
electro-weak Standard Model). The \nc \Uo theory in many aspects behaves like a
non-Abelian gauge theory whose group structure constants depend on the momenta of the
 particles \cite{{Filk},{Shren},{Thomas}}. So, one expects to see the charge quantization
emerging also in the NCQED. In fact this has been shown by Hayakawa \cite{Haya}: the \nc
fermions can carry
charge +1 for $\psi$-type fields, $-$1 for $\chi$-type fields and zero for $\phi$-type
fields ($\phi\to \phi'=U\star\phi\star U^{-1}$)\cite{{Carmelo},{Alvarez}}. We would like
to mention that the latter ($\phi$-type field), although is not carrying any \Uo charge,
similarly to \nc photons, carries the corresponding {\it dipole} moment
\cite{{Shren},{CPT},{Alvarez}}.

\begin{center}
{\bf B. The case with more than one group factor}
\end{center}

So far we have only discussed the gauge groups which were consisting of a simple \nc group
\Un. However,  for building a
physical model, it is necessary and important to consider \nc groups which are semi-simple,
i.e. composed of
some simple \Un\ group factors. To study these cases we have to develop the direct product
of groups in the \nc case. In order to show the obstacle, let us first review the
direct product of groups in the \com case. Let $G_1$ and $G_2$ be two local gauge 
groups. 
Then, the group 
$G=G_1 \times G_2$ is defined through the relations:
\bn\label{direct}
&g&= g_1 \times g_2 \ ,  \ g'=g'_1 \times g'_2,\ \     
g_i, g'_i\in G_i,\ g,g'\in G, \cr 
&g&\cdot g' = (g_1 \times g_2) \cdot ( g'_1 \times g'_2)\equiv 
(g_1 \cdot g'_1) \times ( g_2 \cdot g'_2)\ ,  
\en
where the "$\cdot$" corresponds to the relevant group multiplication. Now, let us turn to
the \nc case and consider $G_1=$\Un\ and $G_2=$\Um. Since both the
\Un\ and \Um\ products, besides the matrix multiplication also involve the \sp, one
cannot re-arrange the group elements and therefore it is not possible to generalize
Eq. (\ref{direct}) to the \nc case. As a consequence of the above argument we cannot have
matter fields which are in fundamental \rep\ of both \Un\ and \Um\ factors. However, still
we
have another possibility left: a matter field, $\Psi$,  can be in the fundamental \rep\ of one
group (e.g. \Un) and {anti-fundamental} \rep\ of the other, i.e.
\be\label{matter}
\Psi\to \Psi'= U\star \Psi\star V^{-1},\ U\in U_{\star}(n), \ V\in
U_{\star}(m).
\ee

For the most general case where the gauge group contains $N$ $U_\star (n_i)$ factors, 
$G=\prod_{i=1}^{N} U_\star(n_i)$, the matter fields can at most be charged under two of
the $U_\star (n_i)$ factors, while they must be singlets under the rest of them. Hence, we
have $N$ types of matter fields which are charged only under one $U_\star (n_i)$ factor  
and ${1\over 2}N(N-1)$ types of them which are carrying two different charges. Therefore, 
altogether we can have ${1\over 2}N(N+1)$ kinds of matter fields.

We would also like to make two other remarks:
\newline
{\it i)} there are also $N$ different $\phi$-type fields, which only carry dipole moments under each
group factors and no net charges.
\newline
{\it ii)} for the gauge bosons we do not face any further problem when the gauge group has more than
one simple \Un\ factor. This is because the gauge fields are always carrying {\it only} one
type of charge (or/and dipole moment), i.e. they are singlets under the remaining group factors.
\vskip .2cm
\begin{center}
{\bf IV. DISCUSSION}
\end{center}
In this letter, elaborating more on the structure of \nc local groups we have uncovered
some facts about these groups and their representations. We show that the closure condition
on the representations of \un\
{\it algebra} restricts these representations only to the one realized through \nbyn\ hermitian
matrices, i.e. higher irreducible representations for the \un\ algebra do not exist.

We have discussed that the concept of a direct product of local gauge groups in the \nc case
cannot be obtained by a simple generalization of the \com case. 
Therefore, the matter fields (which are, in general,  in the fundamental \rep\ of the
group(s)) cannot carry more than two different charges. More explicitly, the matter
fields are either non-singlet under {\it only} one of the simple \Un\ factors of the
semi-simple gauge group or they are in fundamental \rep\ of one factor, while in
anti-fundamental \rep\ of another factor, as indicated in (\ref{matter}).


Although in our group theoretical arguments we have considered the \sp\ (\ref{star}), our
discussions and results are independent of the specific form of the space-time \ncy and hold
for a general \nc product between functions.

We would like to note that, as we have discussed, although \Un\ as a 
\nc group is a {\it simple} one, it still has some sub-groups. These 
sub-groups and their classification in their own turn are very
interesting and important. As the first example, it is straightforward
to check that, for $\lambda\in$\un, ${\rm Tr}\lambda$ forms a \uo 
sub-algebra of \un, and in the same line, the star-exponentiation
of ${\rm Tr}\lambda$ defines a \Uo sub-group. 
Besides this \Uo sub-group which is generated by the trace, \Un\ contains
other sub-groups \Um, $m<n$, of matrices of the form
$\left(\begin{array}{cc}a&0\\0&\ {\bf 1}_{n-m}\end{array}\right)$, $a$ - unitary
$m\times m$ matrix. These sub-groups, and in particular the
trace-generated \Uo\ sub-group discussed above, are needed for 
matter fields which are charged only under a sub-group of \Un, and 
not the whole group. It turns out that such fields are indeed 
necessary for building physical \nc models \cite{NCSM}. As for 
other examples of \Un\ sub-groups, one can define $O_{\star}(n)$ and
$Usp_{\star}(2n)$ \cite{Bars}.

In the present work we have mostly focused on the general properties of the \nc
gauge theories. As for concrete physical models,  the construction of a realistic \nc
version of the Standard Model, i.e. the $SU_c(3)\times SU_L(2)\times U(1) $
gauge theory together with its specific matter content,
which has not the problem of charge $e=0, \pm 1$ quantization, can be the main goal.
Such a theory can be constructed uniquely thanks to the present no-go theorem \cite{NCSM}.


We would like to thank C. Montonen and L. Bonora for helpful discussions and enlightening remarks. 
The financial support of the Academy of Finland under the Project No. 163394
is greatly acknowledged.
The work of P.P. was partially supported by VEGA project 1/7069/20.


\end{multicols}
\end{document}